\documentclass[aps,pre,twocolumn]{revtex4}
\usepackage{amsmath,bm,epsfig}
\newcommand{\ud}{\mathrm{d}}

\newcommand{\beq}{\begin{equation}}
\newcommand{\eeq} [1] {\label{#1} \end{equation}}
\let \= \equiv \let\*\cdot \let\~\widetilde \let\^\widehat \let\-\overline

\def\<{\left\langle}    \def\>{\right\rangle}
\def\({\left(}          \def\){\right)}
\def \[ {\left [} \def \] {\right ]}

\begin{document}
\title{Randomness-Induced Redistribution of Vibrational Frequencies in Amorphous Solids}
\author{Valery Ilyin$^1$, Itamar Procaccia$^1$, Ido Regev$^1$ and Yair Shokef$^2$}
\affiliation{$^1$Department of Chemical Physics, $^2$ Department of Physics of Complex Systems, The Weizmann Institute of Science, Rehovot 76100, Israel}
\
\date{\today}

\begin{abstract}
Much of the discussion in the literature of the low frequency part of the density of states of amorphous solids was dominated for years by comparing measured or simulated density of states to the classical Debye model. Since this model is hardly appropriate for the materials at hand, this created some amount of confusion regarding the existence and universality of the so- called ``Boson Peak'' which results from such comparisons. We propose that one should pay attention to the different roles played by different aspects of disorder, the first being disorder in the interaction strengths, the second positional disorder, and the third coordination disorder. These have different effects on the low-frequency part of the density of states. We examine the density of states of a number of tractable models in one and two dimensions, and reach a clearer picture of the softening and redistribution of frequencies in such materials. We discuss the effects of disorder on the elastic moduli and the relation of the latter to frequency softening, reaching the final conclusion that the Boson peak is not universal at all.
\end{abstract}
\maketitle
\section{Introduction}
\label{intro}

The study of the density of states of solid materials started
with attempts to understand the temperature dependence of the specific heat at
low temperatures, say $C_V\equiv (\partial U/\partial T)_V$ where $U$ is the energy and $T$ the temperature of the system. This called for a microscopic theory for solids,
and the  first one was developed by Einstein, assuming that
in $d$ dimensions each atom is represented as a $d$-dimensional
harmonic oscillator \cite{ein7} (in the original paper the case $d=3$ was
considered). In this article Planck's quantization
assumption, which was originally applied to radiation, was extended to solid vibrations \cite{Kl65}.
In the case of $dN$ linear oscillators each with its own frequency $\omega_i$, Einstein's result can be expressed as
\begin{equation}
C_{V}=dNk_{B}\int\limits_{0}^\infty
\Bigg( \frac{\hbar \omega}{2k_{B}T}\Bigg)^2
csch^2\left(\frac{\hbar \omega}{2k_{B}T} \right) g(\omega) \ud \omega \ .
\label{ein2}
\end{equation}
Here $k_{B}$ and $\hbar$ are  Boltzmann's and Planck's constants respectively, and the
density of states $g(\omega)$ is defined by
\begin{equation}
g(\omega)=\frac{1}{dN}\sum\limits_{i}^{dN}\delta(\omega-\omega_{i})
\label{ein3}
\end{equation}
where $\delta(x)$ is the delta function.

Both the theoretical calculation and experimental measurement of $g(\omega)$ attracted enormous attention over the last century. We are interested in amorphous system like glasses, gels, foams etc., in which randomness appears to influence the low-frequency part of the density of states $g(\omega)$. In particular the relation between the low-frequency behavior and the low-temperature thermodynamics of such systems is of great interest. Studies of the low-frequency part of the density of states are dominated by dividing $g(\omega)$ by the prediction of the Debye model, focusing on the deviation between the two, and in particular on the so-called ``Boson peak'' which emerges in many cases. There exist numerous claims about the Boson peak, its universality \cite{03GPS} and its relation to softening or hardening of the materials under changes of material parameters \cite{08ST}. In this paper we first explain the classical approaches to the issue, including the Debye model and beyond, and then we examine the issue of universality of the Boson peak in one and two dimensions using tractable models that can be computed to desired accuracy. We conclude that there is nothing universal about the Boson peak, and that different types of disorder result in very different redistributions of the low-frequency modes over the spectral domain. There is in general no correlation between the size or the position of the Boson peak and the increase or decrease of elastic moduli.

The structure of this paper is as follows: In Section \ref{Debye_and_Boson} we review Debye's theory and the historical origins of the Boson peak. In Section \ref{beyond} we remind the reader what is entailed in computing the density of states by solving the appropriate eigenvalue equation. We then remind the reader that even for a perfect cubic crystal the Debye model is not exact, with corrections at frequencies which become lower as the material gets softer. To understand the effect of disorder in the inter-particle forces we review some known results and present some new results for one-dimensional chains in Section \ref{oned}. In Section \ref{twod} we discuss tractable models of disorder in two dimensions, aiming to better model the typical disorder exhibited by glass-forming systems. We first examine the effects of disorder in the spring-constants or, equivalently, in the positions of the particles. Second, we consider disorder in the coordination numbers (the number of nearest neighbors), demonstrating that this can lead to major corrections to the Debye form and to very large Boson peaks. This is in general agreement with the idea that the scenario of glass-formation can be encoded by the changing coordination numbers as a function of temperature (so-called upscaling  \cite{scenario1,scenario2,scenario3,scenario4,scenario5,scenario6,scenario7}). In Section \ref{moduli} the elastic moduli and the Debye frequencies which define the `Debye point' are discussed. In Section \ref{discussion} we offer a summary of the paper and some concluding remarks.

\section{Debye's Model and the Boson Peak}
\label{Debye_and_Boson}
\subsection{Debye's Model}
For the sake of simplicity in comparing with
experimental results, Einstein employed a minimal model in which all the oscillators have
the same frequency $\omega_{E}$ \cite{ein7}. Then $g(\omega)=\delta(\omega-\omega_{E})$ and the specific heat is
given by:
\begin{equation}
C_{V}=dNk_{B}\Bigg( \frac{\theta_{E}}{2T}\Bigg)^2
csch^2\frac{ \theta_{E}}{2T},
\label{ein5}
\end{equation}
where $\theta_{E}\equiv \hbar \omega_E/k_B$ is the so-called Einstein temperature.  This minimal model
agreed qualitatively with experimental observations of the specific
heat; nevertheless careful measurements show that the low temperature behavior of
(\ref{ein5}) for a three-dimensional solid, i.e. $C_{V}\sim
3Nk_{B}(\theta_{E}/T)^2exp(-\theta_{E}/T)$  falls off faster than experimental
values. To explain the observed data one needs to account for more adequate vibrational
spectra of actual solids. Debye was the first to link the oscillator frequencies in (\ref{ein2}) with the collective vibrations
of the solid. He treated a solid as an elastic isotropic
 continuum and estimated the density of vibrational states for the case of a
spherical body \cite{D12}. Later it was shown that the frequency distribution
is independent of the shape if the size of a body is large enough and  the surface
contribution can be neglected \cite{MMW63}.  The modern simple derivation of
the Debye distribution can be found, e.g., in \cite{LLst}. The main statement
of the continuum approximation is the linear dispersion relation between the
frequency and the absolute value of the wave vector $k$, $\omega=u_{s}k$ where
$u_{s}$ is the sound velocity. In an isotropic body there exist one
longitudinal and $d-1$ transverse sound waves, therefore, for spatial
dimension $d>1$ one needs two dispersion relations. The Debye density of
states is given by:
\begin{equation}
g(\omega)=\left\{
\begin{array}{ll}
\frac{d}{\omega_{D}}(\frac{\omega}{\omega_{D}})^{d-1}
&\textrm{, if $\omega\le\omega_{D}$} \\
0 &\textrm{, if $\omega>\omega_{D},$}
\end{array} \right.
\label{d1}
\end{equation}
where $\omega_{D}$ is the Debye frequency which defines the cut-off frequency in the spectrum:
\begin{equation}
\frac{1}{(\omega_{D})^d}=\frac{\Omega_{d}}{(2\pi)^d\rho d}
\Bigg(\frac{1}{(u_{l})^d}+\frac{d-1}{(u_{t})^d}\Bigg).
\label{d2}
\end{equation}
Here $\Omega_{d}=(\pi^{d/2})/\Gamma(1+d/2)$ is the coefficient in the
volume definition of $d$-dimensional hypersphere of radius
$r$,  $V_{d}=\Omega_{d}r^d$,
$\Gamma(x)$ is the gamma function,
$\rho$ is the particle number density,
$u_{l}$ and $u_{t}$ are the speeds of
propagation of longitudinal and transverse sound waves:
\begin{equation}
\label{wave}
u_{l}=\sqrt{\frac{dK+2(d-1)\mu}{dm\rho}}; \quad u_{t}=\sqrt{\frac{\mu}{m\rho}},
\end{equation}
where $K$ and $\mu$ are the bulk and shear moduli respectively and $m$ is the molecular mass.

Debye's model has the advantage of a simple analytical form depending
on the elastic properties of a solid only. Due to the long-wavelength approximation
it is insensitive to the microscopic structure. Nevertheless experimental measurements
indicated from the start that Debye's model is far from being the end of the story.

\subsection{The Boson Peak}
The vibrational properties of solids can be investigated
experimentally by studying the inelastic interactions of external radiation
with the solid vibrations. For inelastic scattering of photons one observes
the Raman effect, discovered by Raman in liquids \cite{raman28} and by Landsberg and
Mandelstam in crystals \cite{LM28}. As a result of
this effect the frequency of the incident photon is either red shifted
(Stokes scattering, with high amplitude) or blue shifted (anti-Stokes scattering, with low amplitude).

In crystals, due to the periodic structure, selection rules give
rise to a discrete set of lines. In amorphous materials these spectral
lines broaden, giving rise to a continuous spectrum. The Raman line shape was related
to the density of states of amorphous materials in \cite{SG70} under some assumptions in the
harmonic approximation. The result can be rewritten for Stokes scattering
in the following form:
\begin{equation}
\frac{I_{exp}(\omega)}
{\omega[n(\omega,T)+1]}
=C(\omega) \frac{g(\omega)}{\omega^2},
\label{bp1}
\end{equation}
where $I_{exp}(\omega)$ is the observed Raman intensity at the frequency
shift equal to $\omega$ and
\begin{equation}
n(\omega)=\frac{1}{exp(-\frac{\hbar\omega}{k_{B}T})-1}
\label{bp2}
\end{equation}
is the Bose distribution function.  The function $C(\omega)$ is an empirical function called ``the average light
vibration coupling constant''. Thus the right-hand side of Eq. (\ref{bp1}) is
independent of temperature,  meaning that the temperature dependence of the
Raman intensity should be compensated by the temperature dependence of the
Bose distribution function. This conclusion was confirmed by experiments (see e.g. \cite{hass69}).
At low frequencies the Raman spectrum has a bump whose amplitude changes with temperature.
Once scaled by the Bose function the data at different temperatures collapse to a temperature-independent peak, which is therefore usually referred to as the `Boson peak'.

Analysis of Raman spectra for different amorphous materials indicates
the existence of a Boson peak \cite{MS86}. Therefore, it was suggested
that Raman spectra indicate some universal features of amorphous systems, independent of the
details of molecular interactions. Unfortunately, experimental results determine only the product of
the density of states and the light-vibrational coupling constant, cf. Eq. (\ref{bp1}).
Under the assumption that the low-frequency density of state is defined by the
Debye model, for three-dimensional systems we have
$g(\omega)/\omega^2=const$.   In this case, and only in this case, Eq. (\ref{bp1}) implies that
the coefficient $C(\omega)$ \cite{MS86} must exhibit the Boson peak. On the other hand if
the Debye model does not apply to the particular material at hand, this conclusion cannot be reached.

Additional light was shed on this problem using inelastic scattering of cold neutrons \cite{MMW63}.
Such experiments indicate that in amorphous solids at small frequencies
the quantity $g(\omega)/\omega^2$ is {\em not} constant, showing an excess in the
vibrational density of states \cite{BND84,ZMNPS89}. The vibrational
density of states defines the temperature dependence of the specific heat (\ref{ein2}) and indeed the
latter quantity also displays an excess at low temperature amorphous solids compared to the prediction of Debye's model. This
again provides evidence for additional contributions to the vibrational density of states \cite{AHP86}.

At present it is clear (see, e.g. \cite{ACCKTV09}) that the Raman coupling
coefficient $C(\omega)$ is a rather complicated monotonic function of frequency.
Therefore, the peak in (\ref{bp1}) is defined by the non-monotonic behavior of
the density of vibrational states at low
frequencies (in the sense of Raman scattering). The term `Boson peak' is transferred from the Raman intensity
to the shape of the vibrational density of states at low frequencies. In other
words, the Boson peak describes the {\em deviation} (excess) from the expected
constancy of
$g(\omega)/\omega^{d-1}$ in the $d$-dimensional Debye model.

Once we define the problem of the Boson peak as equivalent to finding the deviations from the Debye model, the
Boson peak is no longer special to amorphous solids. Debye's model takes into account only homogeneous elastic effect; after all, it is well known that the
vibrational spectra of crystals have maxima at the van Hove singularity
points and these maxima are independent of the temperature. The spectral properties in these
regions are  defined only by the lattice structure and the dimensionality
(see, e.g., the well-known exact solution Eq. (\ref{dstatone}) below).

Disorder brings about additional deviations from the Debye model and different kinds of disorder have different effects on the density of states. We will show below that in one-dimensional systems disorder of the inter-particle interactions induces a frequency redistribution with smoothing of the van Hove
singularity. The peak of the spectrum moves to low frequencies with a shift
which depends on the distribution of interactions. The same results
were obtained by direct solution of Eq. (\ref{eigw}) below for three-dimensional
cubic lattices with spring constants distributed in accordance with
a Gaussian \cite{SDG98} or other distributions \cite{KSB01} (in
contrast to one-dimensional chains, three-dimensional cubic lattices are stable
even if some of the spring constants are zero or negative). These and other results using the coherent potential
approximation \cite{ziman} lead to a conclusion that the `Boson peak' in
disordered systems is associated with the lowest van Hove singularity in the spectrum
of the reference crystal \cite{TLNE01}.

It is important to stress that all these results can be taken only as a general indication for the appearance of the `Boson peak' in amorphous solids in two or three dimensions. In all these models only nearest neighbor harmonic interactions (spring constants) were considered. For cubic lattices in two and three dimensions such interaction cannot give rise to a shear modulus, and only the bulk modulus is non-zero. Next-nearest-neighbor interaction are necessary for having a non-zero shear modulus.

\section{Beyond the Debye model: a fair warning}
\label{beyond}

A more general microscopic model of vibrations in a solid was proposed by
Born and von K\'arm\'an \cite{bornkarm12}. In the frame of this model it is assumed
that all the atoms in a crystal interact with spring-like forces and that they vibrate
near fixed equilibrium positions. This harmonic
approximation can be used  both for crystals and amorphous solids;
however, an analytical solution can be obtained only for very simple cases of regular crystal structures.

The total potential energy of a particle configuration
$R=\{ \vec{r}_{1},\vec{r}_{2}, \dots ,\vec{r}_{N} \}$ is
expressed in a pairwise approximation as a sum over pair potentials:
\beq
U_R=\frac{1}{2}\sum\limits_{i\ne j} \phi(r_{ij}).
\eeq{en1}
Relative particle positions are given by vectors
$\vec{r}_{ij}=\vec{r}_{j}-\vec{r}_{i}$, the distance between the $i$th and $j$th
particles is $r_{ij}=|\vec{r}_{ij}|$. Small displacements of all particles $\vec{r}_{i}\to \vec{r^{\prime}}_{i}=\vec{r}_{i}+\delta \vec{r}_{i}$
lead to a new configuration $R^{\prime}=\{
\vec{r^{\prime}}_{1},\vec{r^{\prime}}_{2}, \dots ,\vec{r^{\prime}}_{N} \}$
with the total potential energy:
\beq
U_{R^\prime}=\frac{1}{2}\sum\limits_{i\ne j}
\phi(\mid \vec{r}_{ij}+\delta \vec{r}_{ij}\mid ),
\eeq{en2}
where:
\beq
\delta \vec{r}_{ij}=\delta \vec{r}_{j}-\delta\vec{r}_{i}.
\eeq{dif}

We use the Taylor expansion
\beq
\phi(|\vec{r}+\delta\vec{r}|)=\phi(r)+(\delta\vec{r}\cdot\nabla)
\phi(r)+\frac{1}{2}(\delta\vec{r}\cdot\nabla)^{2}\phi(r)+\ldots
\eeq{taylor}
to obtain
\beq
U_{{  R}^{\prime}}=U_{{ R}}+
\frac{1}{2}\sum\limits_{i\ne j}
\frac{\phi^{\prime}(r_{ij})}{r_{ij}} \vec{r}_{ij}\cdot\delta\vec{r}_{ij}+
\frac{1}{4}\sum\limits_{i\ne j}
\delta \vec{r}_{ij}\cdot  \hat{\mathcal{T}}_{ij} \cdot\delta \vec{r}_{ij},
\eeq{en3}
where
\beq
\hat{\mathcal{{T}}}_{ij}=
\Bigg(\phi^{\prime\prime}(r_{ij})-\frac{\phi^{\prime}(r_{ij})}{r_{ij}}\Bigg)
\vec{n}_{ij}\otimes\vec{n}_{ij}
+\frac{\phi^{\prime}(r_{ij})}{r_{ij}}\mathcal{I}
\eeq{tens1}
is a symmetric tensor $\hat{\mathcal{{T}}}_{ij}=\hat{\mathcal{{T}}}_{ji}$,
$\vec{n}_{ij}=\vec{r}_{ij}/r_{ij}$, and $\mathcal{I}$ is the identity tensor.

Substitution of (\ref{dif}) to (\ref{en3}) yields the dependence the energy of
a harmonic system on particle displacements:
\beq
U_{{ R}^{\prime}}=U_{{ R}}-
\sum\limits_{i}\vec{\mathcal{F}}_{i}\cdot\delta \vec{r}_{i}+
\frac{1}{2}\sum\limits_{i,j}\delta\vec{r}_{j}\cdot
\hat{\mathcal{D}}_{ij} \cdot\delta \vec{r}_{i}.
\eeq{enf}
Here the force applied to the $i$th particle is defined by:
\beq
\vec{\mathcal{F}}_{i}=\sum\limits_{j\ne i}\phi^{\prime}(r_{ij}) \vec{n}_{ij}
\eeq{force}
and the dynamical matrix is given by
\beq
\hat{\mathcal{D}}_{ij}=\left\{
\begin{array}{ll}
\sum\limits_{k\ne i}\hat{\mathcal{T}}_{ik}&\textrm{, if $i=j$} \\
- \hat{\mathcal{T}}_{ij} &\textrm{, if $i\ne j$}
\end{array} \right.
\eeq{u2f}

The equations of motion for a harmonic solid follow from (\ref{enf}):
\beq
m_{i}\frac{\ud^2 \delta {r}_{i}^{\alpha}}{\ud t^2}=
{\mathcal{F}}_{i}^{\alpha}-\sum\limits_{j,\beta}
\hat{\mathcal{D}}_{ij}^{\alpha\beta} \cdot\delta {r}_{i}^{\beta},
\eeq{emovg}
where $m_{i}$ is the mass of $i$th particle and $1\le\alpha,\beta\le d$. In
equilibrium ${\mathcal{F}}_{i}^{\alpha}=0$ and these equations are simplified.

Substitution of a particle displacement of the form
$\delta {r}_{i}^{\alpha}=u_{i}^{\alpha} exp(-i\omega t)$ reduces Eqs.
(\ref{emovg}) to the eigenvalue problem:
\beq
\omega^2u_{i}^{\alpha}=\frac{1}{m_{i}}\sum\limits_{j,\beta}
\hat{\mathcal{D}}_{ij}^{\alpha\beta} \cdot u_{j}^{\beta}
\eeq{eigw}
This equation can be solved directly by diagonalizing $\hat D^{\alpha\beta}_{ij}$ for a system of $N$ particles when $N$ is not too large.
Binning the resulting eigenvalues leads to a histogram that approximates the density of states.
The simplest example is the $d$-dimensional cubic lattice with unstressed distance $a$ between adjacent lattice
points at zero pressure. In the approximation of nearest-neighbor interactions with spring
constants $\phi^{\prime\prime}(a)=\gamma$, the matrix (\ref{tens1}) is given by:
\begin{equation}
\hat{\mathcal{{T}}}^{\alpha\alpha}_{l^{\alpha}m^{\alpha}}=\left\{
\begin{array}{ll}
\gamma&\textrm{, if $\mid l^{\alpha}-m^{\alpha}\mid=1$} \\
0&\textrm{, otherwise,}
\end{array} \right.
\label{tensc}
\end{equation}
where a particle position is defined by the $d$-dimensional vector $\vec{l}a$  with
components $\{ l^{\alpha}a\}$ where $l^{\alpha}$ are integer numbers. For this case the density of states can be found analytically \cite{MMW63}
in the form of an inverse Laplace transform,
\beq
g(\omega)=2\omega \frac{1}{2\pi i}\int\limits_{\sigma-i\infty}
^{\sigma+i\infty}
e^{\omega^2 s} F(s) \ud s,
\eeq{a4}
where the image function $F(s)$ is:
\beq
F(s)=\frac{1}{d}e^{-2 d\tilde{\gamma} s}I_{0}^{d}(2\tilde{\gamma} s).
\eeq{a9}
Here $\tilde{\gamma}=\gamma/m$ and $I_{0}^{d}$ is the modified Bessel function of order zero \cite{AS}.
The inverse Laplace transform in closed analytical form is defined for one and
two dimensional systems, for example if $d=1$ the density of states is given
by:
\beq
g(\omega)=\left\{
\begin{array}{ll}
\frac{2}{\pi}\frac{1}{\sqrt{\omega_{max}^2-\omega^2}}&
\textrm{,  $ 0\leq \omega\leq \omega_{max}$} \\
0&\textrm{, $\omega > \omega_{max}$}
\end{array} \right.
\eeq{dstatone}
where $\omega_{max}=2\sqrt{\tilde{\gamma}}$.
This function diverges at $\omega=\omega_{max}$. This is a general property
of the density of the vibrational states; for {\it periodic structures} there
are integrable singularities (called van Hove singularities) of $g(\omega)$ ($d=1,2$)
or its derivatives ($d=3$). The positions and types of van Hove singularities
depend on the spatial dimension and the topological properties of the crystal
 \cite{MMW63,hove53}. The low frequency behavior of this density of states was computed in \cite{MMW63} with
 the final result
\beq
g(\omega)=\frac{2}{d^2\Gamma(d/2)}\frac{\omega^{d-1}}{(4\pi\tilde{\gamma})^{d/2}}\Bigg(
1+\frac{1}{8\tilde{\gamma}}\omega^{2}+\ldots\Bigg)
\eeq{a16}
The comparison of (\ref{a16}) with the Debye result (\ref{d1}) shows that
{\em the Debye model gives the first term in a more general expansion}. The softer
the system is, the larger is the correction. Substitution of Eq. (\ref{a16}) to
Eq. (\ref{ein2}) allows to estimate corrections to Debye's specific heat
(see, e.g., \cite{leig48}). We note that even for a perfect cubic crystal the Debye model
is not exact, and there can be significant deviations. Clearly when the crystal is not perfect
or when disorder sets in the changes from the Debye limit can become much larger. Thus a blind
comparison of any given density of states to the Debye limit may be unwarranted and can lead to
spurious conclusions. We will come back to this issue when we discuss the Boson peak below.

\section{One-dimensional disordered chains}
\label{oned}
Consider a one-dimensional harmonic chain with lattice spacing $a$ and with random masses and spring
constants as the simplest model of a disordered solid. In the
nearest-neighbor approximation the interaction potential $\phi(r_{ij})$ is
defined by:
\begin{equation}
\phi(r_{i,i+1})=\frac{1}{2}\gamma_{i,i+1}(r_{i,i+1}-a)^2.
\label{enone}
\end{equation}
where $\gamma_{i,i+1}$ are random spring constants taken from a prescribed distribution $p(\gamma)$.
In this case  the definition (\ref{tens1}) reads:
\begin{equation}
\hat{\mathcal{{T}}}_{ij}^{11}=\left\{
\begin{array}{ll}
\gamma_{ij}&\textrm{, if $\mid i-j\mid=1$} \\
0&\textrm{, otherwise}
\end{array} \right.
\label{tone}
\end{equation}
and Eqs. (\ref{eigw}) are  written as:
\begin{eqnarray}
m_{i} \omega^2 u_{i}&=&-\gamma_{i-1.i} u_{i-1}+ \nonumber \\
& &(\gamma_{i-1,i}+\gamma_{i,i+1}) u_{i}-
\gamma_{i,i+1} u_{i+1}.
\label{eigwone}
\end{eqnarray}
Unfortunately, it is impossible to derive a dispersion relations from these
equations  and the analytical solution discussed above becomes
meaningless. Nevertheless, the response of the system to an applied static
force $\Delta P$ can be inferred from the equilibrium conditions which follow
from (\ref{enone}):
\begin{equation}
\gamma_{i,i+1}(r_{i,i+1}-a)=\Delta P.
\label{lstat}
\end{equation}
Summing Eqs. (\ref{lstat}) yields the elongation of the chain:
\begin{eqnarray}
\Delta L&=&\sum\limits_{i}^{N-1}r_{i,i+1}-(N-1)a \nonumber \\
&=&\Delta P\sum\limits_{i}^{N-1}\frac{1}{\gamma_{i,i+1}}.
\label{cone1}
\end{eqnarray}
It is suitable to introduce a quantity
$\frac{1}{ \gamma_{av}}=\langle \frac{1}{\gamma}\rangle$, then the bulk modulus
defined by the condition (\ref{cone1}) is given by:
\begin{equation}
K=\frac{\gamma_{av}}{\rho}.
\label{dkbulk}
\end{equation}
We reiterate that $\gamma_{av}$ is the harmonic average of $\gamma$.
Substitution of (\ref{dkbulk}) to (\ref{wave}) and to (\ref{d2}) yields the following
Debye frequency:
\begin{equation}
\omega_{D}=\pi\sqrt{\frac{\gamma_{av}}{m}} \ .
\label{dfdis}
\end{equation}
If $\gamma_{av}>0$ the Debye frequency has a finite value. Since the Debye model takes into account only the elastic
properties of the material, it should be exact in the
limit $\omega\to 0$ independently of the detailed structure of the material. In this limit every material is an elastic medium. Thus we expect $\lim_{\omega\to 0} g(\omega) =d/\omega_D$ in agreement with the general law (\ref{d1}). We refer to this limit as the ``Debye point''.

In the case $\gamma_{av}=0$ the low frequency behavior of the density of
states depends on the properties of the probability distribution function
$p(\gamma)$ for spring constants \cite{shlomo}.  If
$p(\gamma)_{\gamma\to 0}\to const$ in contrast to the Debye model (\ref{d1})
the density of states exhibits the singular behavior $g(\omega)_{\omega\to 0}\sim
\sqrt{-ln\omega}$.  The density of states in the whole frequency region was
obtained by Dyson in \cite{dyson} analytically for a particular distribution of the ratio of
the spring constants to the masses.  Dyson
introduced a set of new constants $\{ \tilde{\gamma}_{n} \}$ defined by:
\beq
\tilde{\gamma}_{2n-1}=\frac{\gamma_{n,n+1}}{m_{n}},\hspace{10mm}
\tilde{\gamma}_{2n}=\frac{\gamma_{n,n+1}}{m_{n+1}},
\eeq{dset}
and derived an analytic solution for the distribution
\beq
p_{n}(\tilde{\gamma} )=\frac{n^n}{\Gamma(n)}\tilde{\gamma}^{n-1}
e^{-n \tilde{\gamma}}.
\eeq{gamma}
The frequencies are measured in units of $\sqrt{\langle \gamma/m\rangle}$,
and $\langle\tilde{\gamma}\rangle=1$.
For asymptotically large $n$,  $p_{n}(\tilde{\gamma})\to\delta(\tilde{\gamma}-1)$ which
corresponds to the crystal state and the solution coincides with
(\ref{dstatone}) \cite{dyson}.

\begin{figure}
\centering
\includegraphics[width=0.40\textwidth]{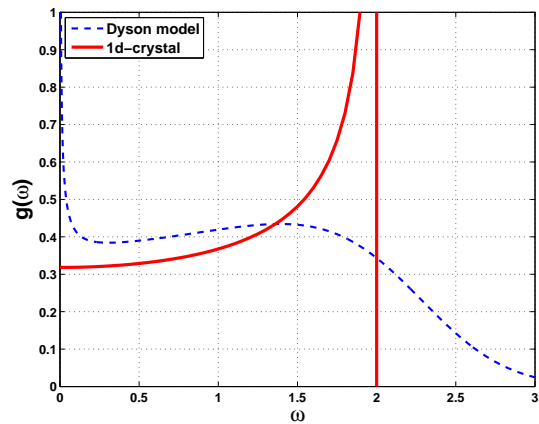}
\caption{Color Online: Density of vibrational states for a one-dimensional crystal (continuous line) and for exponentially distributed random interaction strength (dashed line).}
\label{fig1}
\end{figure}

The density of states for the a special case of the distribution (\ref{gamma}) with
$n=1$ (exponential distribution) is shown in Fig. \ref{fig1}. It is known that
disorder leads to smoothing out any van Hove singularity \cite{ziman}.
The Dyson solution shows that due to the smoothing out of the peak, states
penetrate into the high-frequency region which is forbidden for the periodic structure.
In the small-frequency regime this function diverges
logarithmically in accordance with  the general result of \cite{shlomo}. The
frequencies are redistributed so that the zero-frequency singularity is
followed by a dip. Such behavior is completely different from that of the Debye model.

Unfortunately, it is impossible to study the crossover from Debye to non-Debye
behavior analytically. Nevertheless the density of states of one-dimensional
disordered systems can be estimated with the help of the efficient numerical
method proposed in \cite{dean64} (extension for higher dimensions
is discussed in \cite{DM60,dean72}). This method allows to  calculate
the number of frequencies less than $\omega$ using properties of a Sturm
sequence \cite{QSS2000}. In the following we present calculations pertaining to chains of $10^7$ particles of identical
mass $m=1$; in order to compare different systems we enforced in all cases
cases $\langle \gamma\rangle=1$. The results are summarized as follows:
\subsection{Uniform distribution.}
The simplest distribution function (used also in \cite{dean64} for chains of
$10^3$ particles) is the uniform distribution:
\begin{equation}
p_{u}(\gamma)=\left\{
\begin{array}{ll}
\frac{1}{2\Delta}&\textrm{, if $1-\Delta \le \gamma\le 1+\Delta$} \\
0&\textrm{, otherwise}
\end{array} \right.
\label{udis}
\end{equation}

\begin{figure}
\centering
\includegraphics[width=0.40\textwidth]{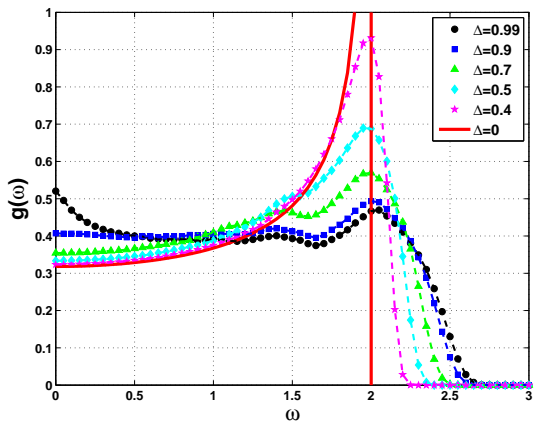}
\caption{Color online: Density of vibrational states for a one-dimensional chain with interactions distributed by the uniform distribution (\ref{udis}). Different symbols pertain to different values of the parameter $\Delta$, see inset.}
\label{udisf}
\end{figure}

For this distribution:
\begin{equation}
\gamma_{av}=\frac{2\Delta}
{ln\frac{1+\Delta}{1-\Delta}}.
\label{avu}
\end{equation}
If $\Delta\to 0$ $p_{u}(\gamma)\to \delta(\gamma-1)$ and the system is reduced
to the homogeneous chain. If $\Delta\to 1$ the spring constant
$\gamma_{av}\to 0$ and one can expect the divergence of $g(\omega)$ at
vanishing frequencies.

The density of vibrational states for the uniform distribution is shown
in Fig.~\ref{udisf}. Upon increasing the parameter $\Delta$ the van Hove
singularity at $\omega=2$ in reduced units is smoothed out and then splits into
two peaks moving in opposite directions. When $\Delta$ approaches unity the
number of low frequency modes increases and a minimum at intermediate
frequencies is formed.

\subsection{Weibull distribution.}
The Weibull distribution is defined by:
\begin{equation}
p_{W}(\gamma)=\frac{\alpha}{\lambda}\Bigg(\frac{\gamma}{\lambda}\Bigg)^{\alpha-1}
e^{-(\frac{\gamma}{\lambda})^\alpha}.
\label{wdis}
\end{equation}
The mean is given by:
\begin{equation}
\langle \gamma\rangle=\lambda \Gamma(1+1/\alpha)
\label{meanw}
\end {equation}
and in order to obtain $\langle \gamma\rangle =1$ the parameter $\lambda$ was set to:
\begin{equation}
\lambda=\frac{1}{\Gamma(1+1/\alpha)}.
\label{lam}
\end{equation}
The average spring constant is given by:
\beq
\gamma_{av}=\frac{\alpha sin(\pi/\alpha)}{\pi}.
\eeq{avw}

\begin{figure}
\centering
\includegraphics[width=0.40\textwidth]{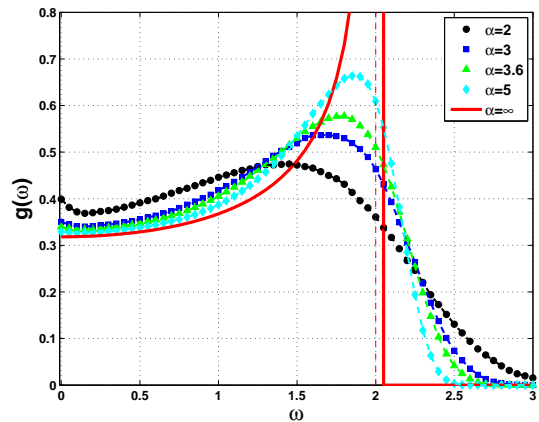}
\caption{Color online: Density of vibrational states for a one-dimensional chain with interactions distributed by the Weibull distribution (\ref{wdis}). Different symbols pertain to different values of the parameter $\alpha$, see inset.}
\label{wwdis}
\end{figure}

When the parameter $\alpha\to \infty$, $p_{W}(\gamma)\to\delta(\gamma-1)$ and
the chain becomes uniform. For $\alpha=1$ the Weibull distribution degenerates
to the exponential distribution and the results obtained in
\cite{dyson} are expected. The  density of the vibrational states for the
Weibull distribution with $\alpha>1$ is shown
in Fig.~\ref{wwdis}. In these cases one peak advances
toward low frequencies, and in the vicinity of zero frequency another peak is
developed.

\subsection{Inverse distribution.}
The exponential distribution of the logarithm of the spring constant was used
in \cite{KSB01} for the investigation of the vibrations of a three dimensional disordered cubic
lattice. This distribution is given by:
\begin{equation}
p_{s}(\gamma)=\left\{
\begin{array}{ll}
\frac{1}{ln\lambda}\frac{1}{\gamma}&\textrm{, if $a \le \gamma\le\lambda a$} \\
0&\textrm{, otherwise}
\end{array} \right.
\label{sdis}
\end{equation}

\begin{figure}
\centering
\includegraphics[width=0.40\textwidth]{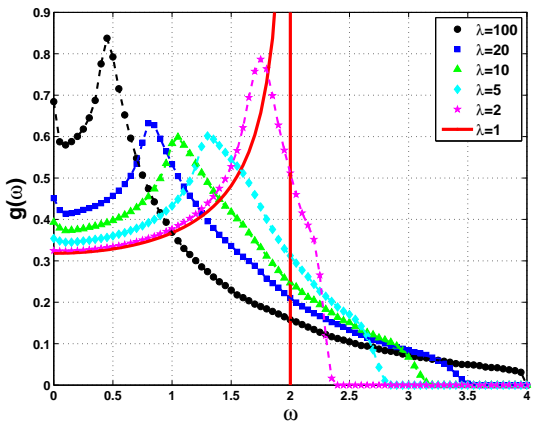}
\caption{Color online: Density of vibrational states for a one-dimensional chain with interactions distributed by the inverse distribution (\ref{sdis}).Different symbols pertain to different values of the parameter $\lambda$, see inset.}
\label{ssdis}
\end{figure}

The mean is $a(\lambda-1)/ln\lambda$, therefore,  the parameter $a$ is
defined by:
\begin{equation}
a=\frac{ln\lambda}{\lambda-1}.
\label{para}
\end{equation}

The average spring constant is given by:
\begin{equation}
\gamma_{av}=\lambda\Bigg(\frac{ln\lambda}{\lambda-1}\Bigg)^2.
\label{avki}
\end{equation}
$\lambda=1$ corresponds to the
ordered chain. The densities of the vibrational states for the distribution
(\ref{sdis}) with different values of $\lambda$ are shown
in Fig.~\ref{ssdis}. For large $\lambda$ a dip at low frequencies develops
and is followed by a peak, both moving to smaller frequencies with
increasing the parameter $\lambda$.
\begin{figure}
\centering
\includegraphics[width=0.40\textwidth]{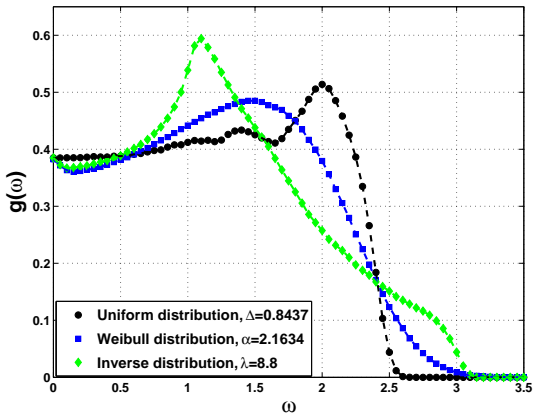}
\caption{Color online: Comparison of the vibrational density of states for a common value $\gamma_{av}=0.6838$.
The symbols are explained in the inset.}
\label{puws}
\end{figure}

Finally, recall that the Debye behavior of a disordered chain is defined by its elastic properties
(\ref{dfdis}), i.e., by $\gamma_{av}$. Therefore it is useful to compare
vibrational properties of different chains with the same bulk modulus, cf.
Fig.~\ref{puws}. Note that this figure is interesting from the point of view of comparing with
Debye's model. The Debye point at $\omega=0$ is the same for all three models since we chose
$\gamma_{av}$ to be the same; hence the Debye frequency (\ref{dfdis}) is identical
for these three models. We could therefore expect identical Debye predictions for
these three models. In contrast, the actual density of states presents widely
different frequency dependence for the three models. This means that the re-distribution
of frequencies depends on the nature of randomness and is not only a function of the elastic
properties.


\section{Two-dimensional systems}
\label{twod}

In an amorphous solid the distance between two particles, their relative orientation, and the number of nearest and next-nearest neighbors of each particle all possess some randomness and thus all the terms in $\hat{\mathcal{{T}}}_{ij}$ are random and so are the elements of the dynamical matrix $\hat{\mathcal{D}}_{ij}$ in Eq. (\ref{eigw}). In order to study the effect of different types of disorder on the spectrum of the dynamical matrix we examine several models of elastic networks.


\subsection{Elastic triangular anti-ferromagnet}

The anti-ferromagnetic Ising model on a rigid triangular lattice is geometrically frustrated since the energy of the three bonds on each triangular plaquette of the lattice may not be simultaneously minimized \cite{Wannier,Houtappel}. This leads to a highly degenerate ground-state and thus to unconventional phases of matter \cite{Moessner01,Ramirez03,Tarjus05,Moessner06}. Allowing the lattice to deform may relieve this frustration and lift the ground-state degeneracy \cite{CK86,Bulbul}. Recent experimental \cite{Yilong} and theoretical \cite{YairTom} studies have shown that frustration is only partially relieved and that such systems exhibit glassy behavior, dramatically slow down, and fall into metastable disordered configurations. In order to allow the system to obtain a disordered deformation, we follow \cite{YairAntonTom} and assume a pair potential for Eq. (\ref{en1}) of the form:
\begin{equation}
\phi_{ij}(r_{ij}) = -J\ \sigma_{i}\sigma_{j}\ [1 - \epsilon\ (r_{ij} - a)]+ \frac{1}{2}\gamma\ (r_{ij} - a)^2
\label{afm},
\end{equation}
Here we assume the spin variables $\sigma_{i} = \pm 1$ are random. The magnetic interaction is taken to be anti-ferromagnetic $J<0$, $\epsilon>0$ controls the magneto-elastic coupling strength, and $\gamma>0$ is the stiffness of the uniform springs connecting each nearest-neighbor pair. Note that some previous studies of the density of vibrational states of elastic networks (see for example \cite{TLNE01}) focused on harmonic lattices with random spring constants (a multidimensional version of the one-dimensional analysis provided in Section \ref{oned}). In the models we use here, the strength of the interaction is randomized by frustration and non-harmonic terms in the potential and therefore arises more naturally; we do not need any assumptions about the distributions that govern the interactions. In this sense the disorder in this model and its derivatives are similar in nature to the disorder in glass forming molecular systems. Another important difference with respect to conventional lattice models is the effect of off-lattice positional disorder on the terms that depend on the relative distance and orientation between two particles in the matrix $\hat{\mathcal{T}}_{ij}$, which in the present case reads
\beq
\hat{\mathcal{{T}}}_{ij}=
\Bigg(\gamma-\frac{J\ \sigma_{i}\sigma_{j}\ \epsilon +\gamma\ (r_{ij} - a)}{r_{ij}}\ \Bigg)
\vec{n}_{ij}\otimes\vec{n}_{ij}\nonumber
\eeq{}
\beq{}
+\frac{J\ \sigma_{i}\sigma_{j}\ \epsilon +\gamma\ (r_{ij} - a)}{r_{ij}}\mathcal{I}
\eeq{}

\begin{figure}
\centering
\includegraphics[width=0.45\textwidth]{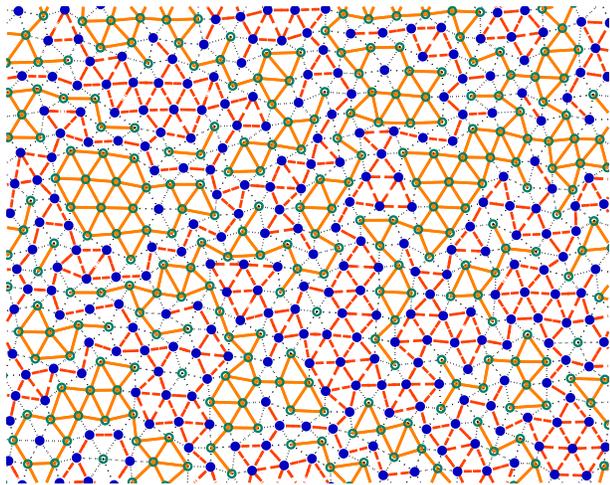}
\caption{Color online: Typical realization of the elastic triangular anti-ferromagnet (\ref{afm}). Solid points represent up spins and empty points represent down spins. dashed lines connect interacting up spins and solid lines connect interacting down spins. Dotted lines connect pairs of interacting up and down spins.}
\label{net}
\end{figure}

We calculated the density of states for $20$ realizations with $6400$ particles each. Each realization of the system was initiated by positioning the particles on a triangular lattice with periodic boundary conditions. Each particle was assigned a random spin value and the energy of the entire network was minimized, using the conjugate-gradient method \cite{mod}. The minimization was achieved by changing the coordinates of the particles, keeping the interaction between the original nearest-neighbors only, and keeping the spin values fixed. A typical resulting configuration is shown in Fig. \ref{net}. The density of states was then calculated using the eigenvalue Eq. (\ref{eigw}). The square-roots of the eigenvalues were collected in bins and the histogram recorded. To compute $g(\omega) /\omega)$ the most precise method turned out to be calculating $g(\omega)/\omega\equiv2 G(\omega^2)$ where $G(\omega^2)$ is the histogram of the eigenvalues themselves. In order to compute $g(\omega)/\omega$ at $\omega=0$ we employed Eq. (\ref{d2}) and the elastic moduli computed below. The same
method was used for all the models listed below. Throughout, we set $\gamma=1$, $m=1$, $J=1$, $a=1$ and measure the density of states for various values of $\epsilon$ and of the other parameters defined for the subsequent models.

\begin{figure}
\centering
\includegraphics[width=0.40\textwidth]{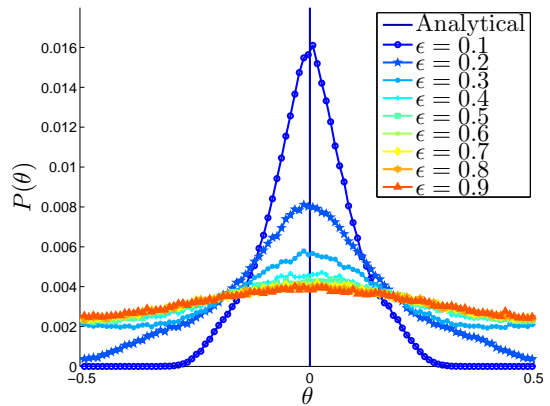}
\caption{Color online: Distribution of angles between nearest-neighbors in the elastic triangular anti-ferromagnet (\ref{afm}) for various values of $\epsilon$, see inset.}
\label{angles}
\end{figure}

Figure \ref{angles} shows the effect of disorder on the distribution of angles $\theta$ between the inter-particle bonds and the $\hat x$ axis. This angle determine the value of the term $\vec{n}_{ij}\otimes\vec{n}_{ij}$. In a perfect triangular lattice these angles take six discrete values $\theta_i= \pi/3i$, ($0\le i\le 5$). In the disordered system these angles have a smooth distribution, and due to isotropy it is sufficient to consider the distribution of of the angles of one bond, say between $[-\pi/6,\pi/6]$.

\begin{figure}
\centering
\includegraphics[width=0.40\textwidth]{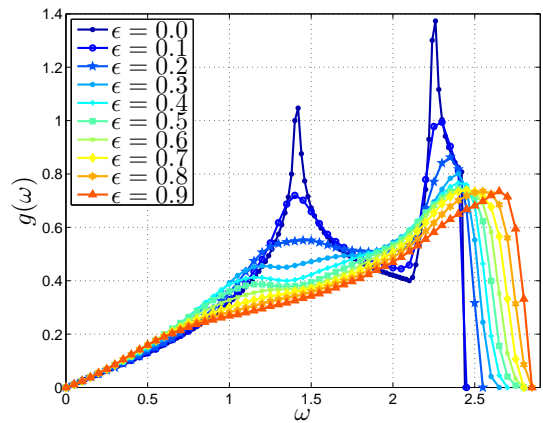}
\caption{Color online: Vibrational density of states in the elastic triangular anti-ferromagnet (\ref{afm}) for various values of $\epsilon$, see inset.}
\label{orig}
\end{figure}

Figure \ref{orig} shows the effect of disorder on the density of states. For the ordered triangular lattice the density of states exhibits van Hove singularities \cite{Dean63}. The most obvious effect of disorder is the smearing of the singularities and the flattening of the density of states. This results in filling the gaps between the singularities but also in some modes leaking to higher and lower frequencies. In particular, there is a change in the density of states at low-frequencies compared to the tail that characterizes the ordered lattice. It is important to note that when $\epsilon$ becomes too large, the network begins to fold upon itself. In a more realistic model, say with next-nearest-neighbor interactions, where the particles are not physically linked to each other this folding is relieved by changing the coordination number (i.e. number of neighbors). Below we will also study the effect of randomizing the coordination number.

\begin{figure}
\centering
\includegraphics[width=0.45\textwidth]{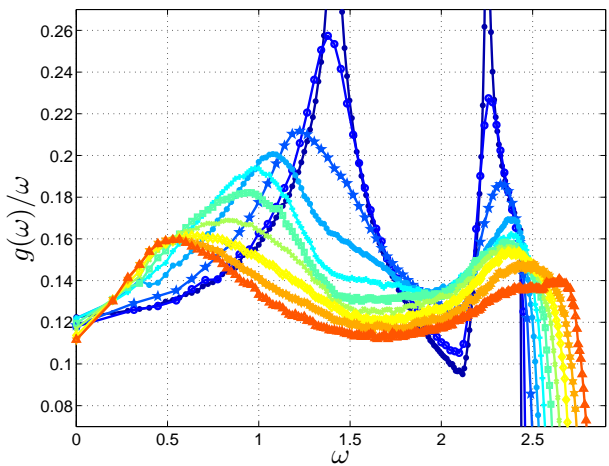}
\caption{Color online: Vibrational density of states normalized by Debye's prediction in the elastic triangular anti-ferromagnet (\ref{afm}). Symbols are the same as in Fig. \ref{orig}}
\label{overw}
\end{figure}

To emphasize the deviation from Debye's model we examine in Fig. \ref{overw} the density of states divided by the prediction of Debye's model, which for $d=2$ is linear in frequency. A peak at low frequencies is observed and its position shifts to lower frequencies as the disorder increases. However, its height decreases upon increasing disorder. Thus {\em in this model there is a negative correlation between the magnitude of disorder and the amount of the deviation from the Debye model} (see also Figs. \ref{udisf} and \ref{wwdis}). Note that in this example it is hard to notice the deviation from the Debye model at low frequencies without dividing the density of states by $\omega$.

Examining Figs. \ref{orig} and \ref{overw} we note that the density of states reaches zero at zero frequency in accordance with (\ref{d1}). Dividing by $\omega$ we observe a finite limit in Fig. \ref{overw}. This behavior follows from Eq. (\ref{d1}) which predicts such a finite limit at $d=2$.


\subsection{Non-Linear Springs}

Here we investigate the effect of random contributions to the harmonic part of the potential. This will
bring us closer to generic systems. There is more than one way of doing so, and we therefore consider two different models for the inter-particle potential. The first has the form
\begin{eqnarray}
\phi_{ij}(r_{ij}) &=& -J\ \sigma_{i}\sigma_{j}\ [1 - \epsilon\ (r_{ij} - a)]  \nonumber\\
&+& \frac{1}{2}\gamma\ (r_{ij} - a)^2 + \frac{1}{3}\kappa\ (r_{ij} - a)^3 \label{unharmonic}
\end{eqnarray}
The harmonic term now reads:
\begin{equation}
\phi''(r_{ij}) = \gamma + 2\kappa(r_{ij} - a)\ .
\end{equation}
Due to the fluctuation in the inter-particle distances around $a$, this term fluctuates around an average value $\gamma$.

\begin{figure}
\centering
\includegraphics[width=0.40\textwidth]{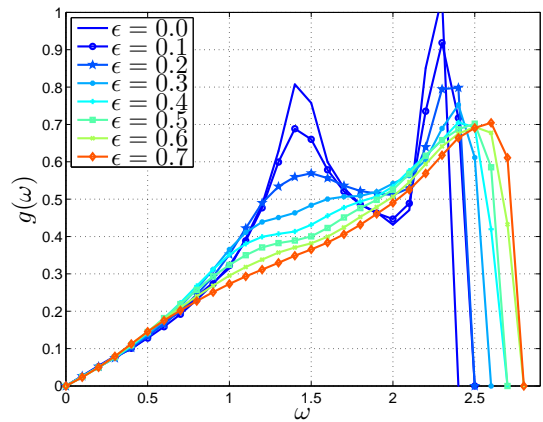}
\caption{Color online: Density of states for the model with non-linear elasticity (\ref{unharmonic}) with $\kappa=0.25$ and different values of $\epsilon$, see inset.}
\label{unhar}
\end{figure}

\begin{figure}
\centering
\includegraphics[width=0.40\textwidth]{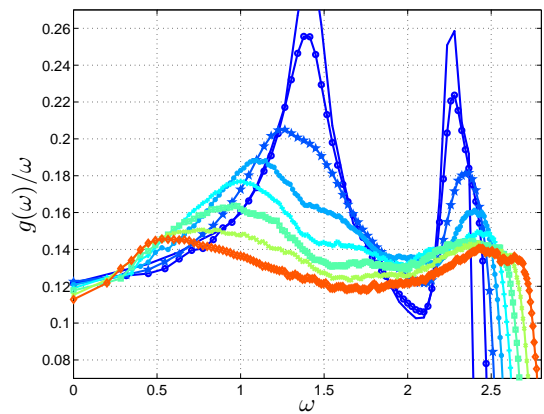}
\caption{Color online: Density of states for the model with non-linear elasticity (\ref{unharmonic}) (Fig. \ref{unhar}) normalized by Debye's model. Symbols are the same as in Fig. \ref{unhar}}
\label{gwunha}
\end{figure}

We repeated the procedure described above for calculating the density of states for $\kappa=0.25, 0.5$ and 1, and for various values of $\epsilon$. We observed the same qualitative behavior for all $\kappa$ values and present in Figs. \ref{unhar} and \ref{gwunha} the raw density of states and the result after normalizing by Debye's prediction for $\kappa=0.25$. As with the first model, we see excess modes at low frequencies


\subsection{Magneto-elastic coupling}

The second way to modify the elastic triangular anti-ferromagnet (\ref{afm}) is by adding a non-linear separation dependence to the magneto-elastic coupling term:
\begin{eqnarray}
\phi_{ij}(r_{ij}) &=& -J\ \sigma_{i}\sigma_{j}\ [1 - \epsilon\ (r_{ij} - a) + \frac{1}{2}\nu(r_{ij} - a)^2 ] \nonumber\\
&+& \frac{1}{2}\gamma\ (r_{ij} - a)^2 \label{exHar}.
\end{eqnarray}
The harmonic term in this case reads:
\begin{equation}
\phi''_{ij}(r_{ij}) = \gamma -J\sigma_i\sigma_j \nu \ .
\end{equation}

\begin{figure}
\centering
\includegraphics[width=0.40\textwidth]{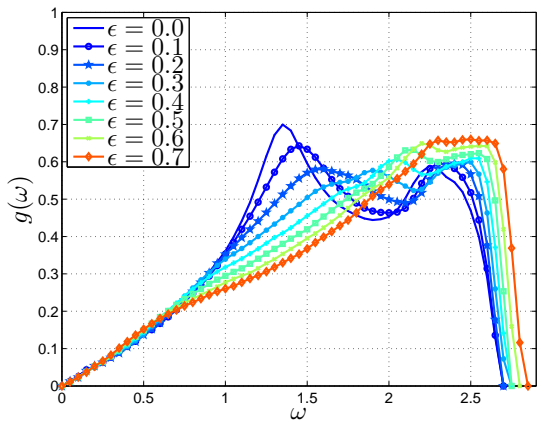}
\caption{Color online: Effect of the non-linear magneto-elastic coupling (\ref{exHar}) on the density of states for $\nu=0.3$ and different $\epsilon$ values, see inset.}
\label{harmspin}
\end{figure}

\begin{figure}
\centering
\includegraphics[width=0.40\textwidth]{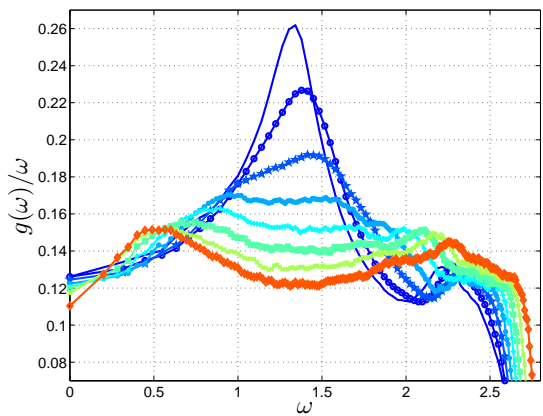}
\caption{Color online: The density of states with non-linear magneto-elastic coupling (Fig. \ref{harmspin}) divided by the Debye behavior. The symbols are the same as in Fig. \ref{harmspin}.}
\label{gwxha}
\end{figure}

The density of states for this case is shown in Figs. \ref{harmspin} and \ref{gwxha} for a representative value of $\nu=0.3$. Qualitatively similar results were obtained for $\nu=0.45$. We recognize in these figures a smoothing of the Van-Hove singularities with redistribution toward both lower and higher frequencies. As before, we see a peak at low frequencies which moves toward lower frequencies when the disorder parameter $\epsilon$ is increased. We will next examine the effect of topological disorder and see that this type of disorder has a much more pronounced effect on the density of the low-frequency states.


\subsection{Randomly diluted elastic triangular anti-ferromagnet}

\begin{figure}
\centering
\includegraphics[width=0.40\textwidth]{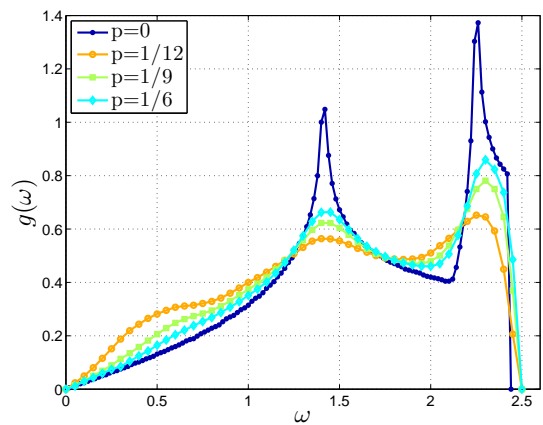}
\caption{Color online: Density of states of the diluted elastic triangular anti-ferromagnet (\ref{topo}) with $\epsilon=0.1$ and various $p$ values, as indicated in the inset.}
\label{Top1}
\end{figure}

\begin{figure}
\centering
\includegraphics[width=0.40\textwidth]{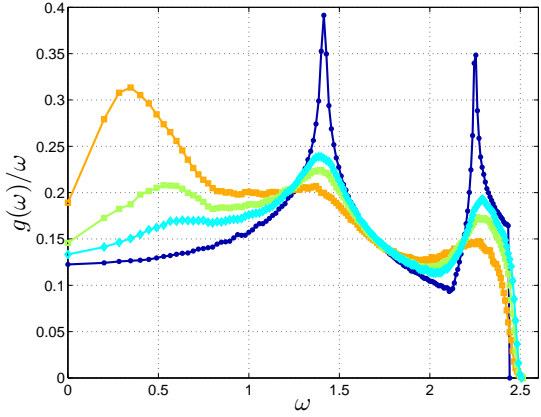}
\caption{Color online: Density of states of the diluted elastic triangular anti-ferromagnet (\ref{topo}) (Fig. \ref{Top1}) normalized by the Debye prediction. Symbols are the same as in Fig. \ref{Top1}.}
\label{Top2}
\end{figure}

It has recently been argued that the glass transition involves a change in the number of neighbors that each particle has \cite{scenario1,scenario2,scenario3,scenario4,scenario5,scenario6,scenario7}. Moreover, recent work on jammed sphere packings has indicated the relevance of the coordination number near isostaticity for determining low-frequency vibrational modes \cite{Silbert05,Wyart05,Xu07}. In order to account for these effects we use the elastic triangular anti-ferromagnet (\ref{afm}) but include the possibility of a missing link between two neighbors:
\begin{equation}
\phi_{ij}(r_{ij}) = g _{ij}\{-J\sigma_{i}\sigma_{j}[1 - \epsilon\ (r_{ij} - a)] + \frac{1}{2}\gamma\  (r_{ij} - a)^2\}\label{topo}
\end{equation}
Where $g_{ij}$ is $0$ with probability $p$ and $1$ with probability $1-p$. We use $p$ close to $0$ to avoid rigidity percolation \cite{FengSen84,GarbocziThorpe85} and keep at least $3$ bonds per particle in order to avoid floppy modes (modes of zero frequencies). We thus create a sparse network with a local coordination number varying between $6$ and $3$. This model is a modification of the model described in \cite{FTG} which studied the effect of disconnecting links of a harmonic triangular lattice. In contrast to that model, our model introduces disorder in the equilibrium positions of the particles as well as in their coordination number. We solved this model as before, by first finding the lattice deformation that locally minimizes the mechanical energy. In this   model the effect of disorder on the low-frequency domain of the density of states is much larger than before, as seen in Figs. \ref{Top1} and \ref{Top2}.

The slope of the density of states at low frequencies, although linear, as expected by Debye's theory, is very different from the slope of the density of states for the perfect lattice (see also Fig. \ref{ssdis}). However this modified Debye point is consistent with the system's elastic moduli, as will be descried in the following section.


\section{Elastic Moduli}
\label{moduli}

To better understand the differences in types of randomness and their effect on the frequency redistribution we consider here the elastic moduli of the two-dimensional models treated above. Contrary to the density of states, the elastic moduli are global measures of the response of the system to external mechanical perturbations. Nevertheless there exists an interesting relation between these global properties and the frequency redistribution.

Measurements of the shear modulus were done by applying affine shear transformations, minimizing the energy after each step using the Lees-Edwards periodic boundary conditions, in order to measure the stress. After each minimization the stress for each particle was measured directly from its microscopic definition and the mean stress was computed as a sum over all particles. Next, the mean stress as a function of the strain was calculated, and the shear modulus was extracted from the numerical derivative. The bulk modulus was measured by decreasing the volume and measuring the diagonal part of the stress tensor (the pressure). The elastic moduli were used to calculate the Debye point at $\omega=0$.

\begin{figure}
\centering
\includegraphics[width=0.40\textwidth]{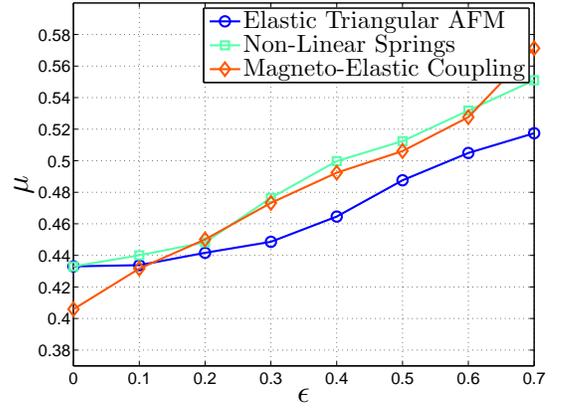}
\caption{Color online: Shear moduli of the three models Eqs. (\ref{afm}) (circles) , (\ref{unharmonic}) (squares) and (\ref{exHar}) (diamonds) for different values of the disorder control parameter $\epsilon$.}
\label{original_shear}
\end{figure}

\begin{figure}
\centering
\includegraphics[width=0.40\textwidth]{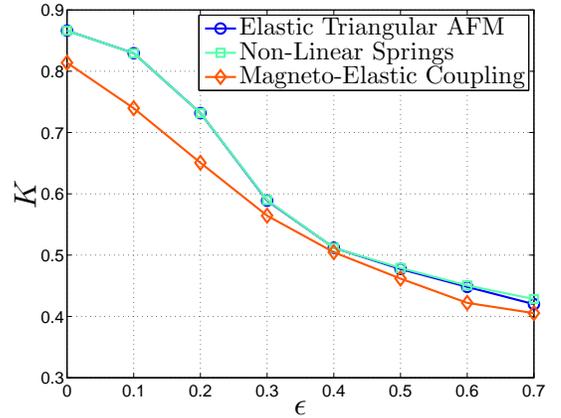}
\caption{Color online: Bulk modulus of the three models Eqs. (\ref{afm}) (circles) , (\ref{unharmonic}) (squares) and (\ref{exHar}) (diamonds) for different values of the disorder control parameter $\epsilon$.}
\label{original_bulk}
\end{figure}

We first measured the elastic moduli for the first three models Eqs. (\ref{afm}), (\ref{unharmonic}) and (\ref{exHar}) for different values of $\epsilon$ (see Figs. \ref{original_shear} and \ref{original_bulk}). The results are somewhat unexpected. In all three models the shear modulus increases when disorder is increased, while the bulk modulus decreases. Thus in these three models we cannot say that the system softens or hardens, since one elastic modulus decreases while the other increases.

\begin{figure}
\centering
\includegraphics[width=0.40\textwidth]{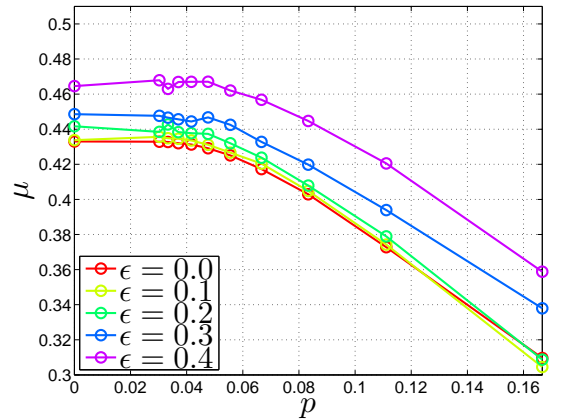}
\caption{Shear modulus of the diluted model Eq. (\ref{topo}) as a function of $p$ and $\epsilon$. The shear modulus increases when $\epsilon$ increase but it decreases when $p$ increases.}
\label{topo1}
\end{figure}

\begin{figure}
\centering
\includegraphics[width=0.40\textwidth]{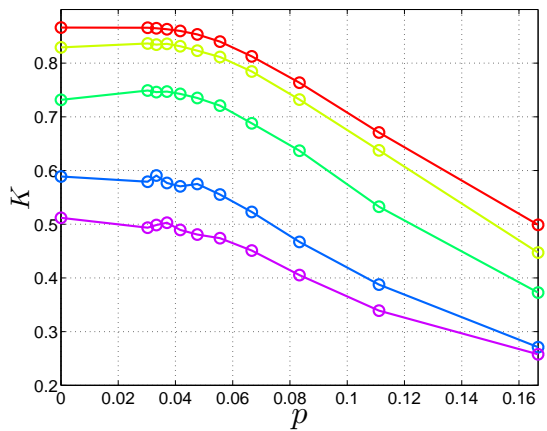}
\caption{Color online: Bulk modulus of the diluted model Eq. (\ref{topo}) as a function of $p$ and $\epsilon$, symbols as in Fig. \ref{topo1}. The bulk modulus decreases when $\epsilon$ increases but it decreases like the shear modulus when $p$ increases.}
\label{topo2}
\end{figure}

For the diluted network (\ref{topo}), both elastic moduli decrease with increasing $p$, see Figs. \ref{topo1} and \ref{topo2}. This is very physical; cutting bonds must result in true softening of the system. Note that for a fixed value of $p$ the qualitative behavior with $\epsilon$ is similar to the previous three models.

\begin{figure}
\centering
\includegraphics[width=0.40\textwidth]{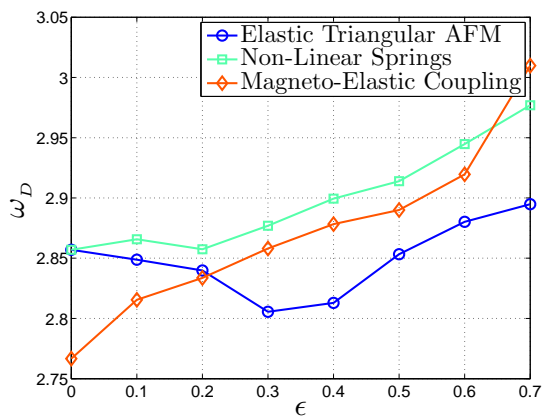}
\caption{The Debye frequency of the three models Eqs. (\ref{afm}) (circles) , (\ref{unharmonic}) (squares) and (\ref{exHar}) (diamonds) for different values of the disorder control parameter $\epsilon$.}
\label{Debye3}
\end{figure}

\begin{figure}
\centering
\includegraphics[width=0.40\textwidth]{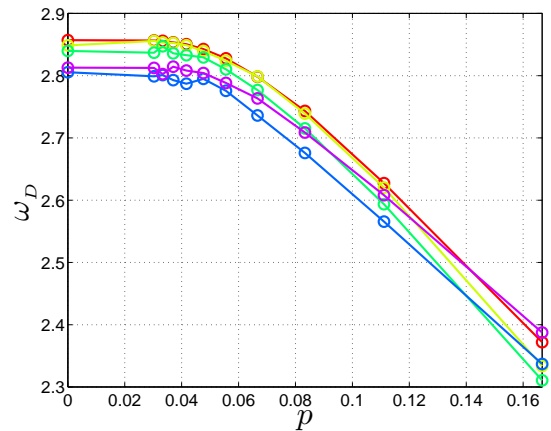}
\caption{Color online: The Debye frequency of the diluted model Eq. (\ref{topo}) as a function of $p$ and $\epsilon$, symbols as in Fig. \ref{topo1}. }
\label{Debyetopo}
\end{figure}

We see that this last model differs from the previous three in having clear softening when the parameter $p$ increases.
One way to take into account both moduli in discussing the softening of the system is by focusing on the Debye point
\begin{equation}
\lim_{\omega \to 0} \frac{g(\omega)}{\omega^{d-1}} = \frac{d}{(\omega_D)^d} \ . \label{Dp}
\end{equation}
We computed the Debye frequency for the four models at hand, and the results are presented for the first three models
in Fig. \ref{Debye3} and for the fourth model in Fig. \ref{Debyetopo}.

We see from Fig. \ref{Debye3} that the Debye frequency is practically constant for the first model, and slightly increases with $\epsilon$ for the second and third models. For the fourth model (Fig. \ref{Debyetopo}) the Debye frequency softens dramatically when the parameter $p$ is changed. The disorder governed by the parameter
$\epsilon$ almost does not change the Debye frequency also in this model.

\section{Discussion and Conclusions}
\label{discussion}

The main conclusion from the one-dimensional and two-dimensional examples treated above are as follows:

\begin{enumerate}

\item Both ordered and amorphous solids exhibit peaks in their density of states. In ordered solids these peaks are understood as van Hove singularities. In amorphous solids these singularities are smoothed out, providing higher amplitudes to both lower and higher frequencies.

\item For both the crystalline and the amorphous examples the comparison with the Debye model shows complete agreement only at $\omega\to 0$, independently of the dimensionality. Within the Debye model we expect $g(\omega)/\omega^{d-1}$ to be constant. This seems to be never the case.

\item Dividing the computed density of states by $\omega^{d-1}$ reveals the so-called ``Boson peak''. Its position and amplitude depend on many details; in one-dimensional cases we showed how it depends on the statistical distribution of the spring constants. In two-dimensions we showed how it depends mainly on the spatial disorder and on the coordination number, with the latter being dominant. In this sense there is nothing universal about the Boson peak. We cannot even tell a-priori whether increasing disorder might increase or decrease the amplitude of the Boson peak, cf. Fig. \ref{last}
\begin{figure}
\centering
\includegraphics[width=0.40\textwidth]{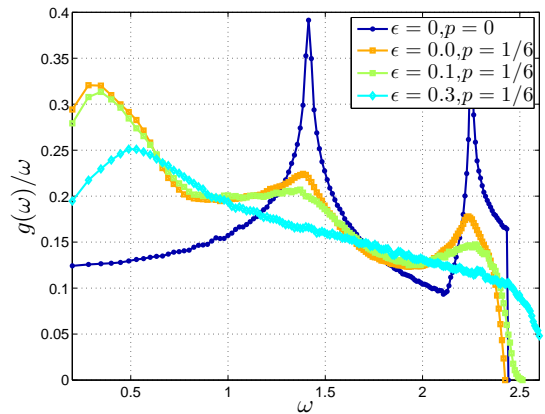}
\caption{Color online: $g(\omega)/\omega)$ for the diluted model with different values of $\epsilon$.
Note that the Boson peak amplitude is reduced when $\epsilon$ which randomizes the particle positions is increased.  }
\label{last}
\end{figure}

\item We cannot discern any clear correlation between the Boson peak and the elastic moduli. In one dimension we showed (Fig. \ref{puws}) that three models with identical bulk modulus exhibit completely different redistributions of frequencies. In two-dimensions we showed for the first three models that the bulk modulus decreased with disorder whereas the shear modulus increased, contrary to expectations.  The change in the Debye frequency is small; nevertheless we have completely different frequency re-distributions. In the fourth two-dimensional model we considered, both elastic moduli decrease simultaneously and we indeed saw a pronounced redistribution to lower frequencies when the average coordination number changed. Again we see no systematic correlation with the behavior of the elastic moduli.

\end{enumerate}

Of course, all these conclusions concern the simple models discussed above. Nevertheless the phenomena discussed are not special to these models or even to amorphous solids in general. An experimental connection between shear modulus and the low-frequency behavior of the vibrational spectrum was given in \cite{SSH06}  by analyzing the low-temperature specific heat. In contrast to the common view that excess in low-temperature specific heat (and, hence, in low-frequency modes) is special to disordered systems only, it was demonstrated that crystals and amorphous solids with almost the same shear modulus have quite different positions of the Boson peak. This is in accord with our conclusions that with the same Debye point we can have different redistributions of frequencies.

In summary, it is quite possible that in a given family of amorphous materials, where the randomness is quite similar, there can be a correspondence between the redistribution of frequencies and the shear modulus. However this is not a general correlation, as we saw with the present examples. We saw that the actual density of states is a complicated function of many competing influences. It is unlikely that one given parameter of whatever nature (like the shear modulus) can capture this full complexity. The understanding of the density of states and its changes under modified interactions remains a theoretical calculation of significant difficulty.

\end{document}